\documentclass[12pt]{iopart}
\usepackage{iopams}
\begin{document}

\title{Covariant formulation of electrodynamics in isotropic media revisited}
\author{D V Red\v zi\' c}

\address{Retired from Faculty of Physics, University of Belgrade, PO
Box 44, 11000 Beograd, Serbia} \eads{\mailto{redzic@ff.bg.ac.rs}}

\begin{abstract}
This paper resolves a persistent ambiguity  regarding the covariant
formulation of electrodynamics in a vacuum, as well as of
Minkowski's electrodynamics of moving isotropic media. By analyzing
a recent debate, we demonstrate that current interpretations
misrepresent the relationships between the electromagnetic
four-vector potential, gauge conditions, and field tensors. We
specifically challenge the notion that the geometric status of
$A^\mu$ as a four-vector is contingent upon the choice of gauge. We
trace these inconsistencies back to potentially misleading
treatments of the four-vector potential in the classic texts of
M{\o}ller, Landau and Lifshitz, Jackson and Griffiths. Beyond
correcting the specific conceptual errors in the recent literature,
this work clarifies the theoretical framework for ensuring the
electromagnetic potential remains consistent with gauge invariance
and the principle of relativity in macroscopic isotropic media.
\end{abstract}


\vspace {5mm}

\section{Introduction}
This paper analyzes a recent series of publications in the
\emph{European Journal of Physics} concerning the covariant
formulation of Minkowski's electrodynamics of moving isotropic
media. We specifically examine the original article by Pal [1], the
subsequent Comment by Franklin [2], and the Reply by Pal [3]. While
Pal's Reply successfully demonstrates that the specific points
raised in Franklin's Comment regarding the validity of the original
work are invalid, we show that all three papers nonetheless  contain
misrepresentations of both Minkowski's electrodynamics and covariant
vacuum electrodynamics. Specifically, both Pal and Franklin
misinterpret the relationships between the four-vector potential,
gauge invariance and field tensors. A possible source of these
errors is identified in the potentially misleading treatments of the
electromagnetic four-vector potential found in several classic and
highly influential texts, specifically those of C. M{\o}ller [4], L.
D.Landau and E. M. Lifshitz [5], J. D. Jackson [6] and D. J.
Griffiths [7]. By dissecting the mathematical inconsistencies
present in the recent debate, we clarify the relationship between
the four-vector potential, gauge invariance, and the field tensors
within the context of moving media. The present work provides the
necessary mathematical corrections and conceptual clarifications to
ensure full consistency  with standard covariant theory and the
principle of relativity.

Section 2 contains a brief review of some basic results of covariant
electrodynamics in the vacuum, and of potentially misleading
treatments of the four-vector potential in the classic texts [4-7].
In Section 3 we discuss Pal's original paper [1] and point out
errors and misinterpretations in Franklin's Comment [2] and Pal's
Reply [3].

\section{Some basic results of covariant electrodynamics in the
vacuum} In this Section we recall some basic results of covariant
electrodynamics in the vacuum which are relevant for the present
discussion. For the convenience of the reader, throughout this paper
we use the notation and units (Heaviside--Lorentz) of Refs. [1] and
[3].

The electromagnetic field tensor $F^{\mu\nu}$ is defined by

\begin{equation}
F^{\mu\nu} = \partial^\mu A^\nu - \partial^\nu A^\mu\, ,
\end{equation}
where $A^\mu$ is by definition the four-vector for electromagnetic
potentials which is the four-component combination of the scalar and
vector potentials $\varphi$ and $\vec A$

\begin{equation}
A^\mu = (\varphi, A_x, A_y, A_z)\, ,
\end{equation}
and

\begin{equation}
\partial_\mu \equiv \frac {\partial}{\partial x^\mu}\, , \qquad x^\mu =
\{ct,x,y,z\}\, ;
\end{equation}
the metric employed is

\begin{equation}
\eta_{\mu \nu} = \mbox{diag} (+1,-1,-1,-1)\, ,
\end{equation}
so that

\begin{equation}
A_\mu = \eta_{\mu \nu}A^\nu = (\varphi, -A_x, -A_y, -A_z)\, .
\end{equation}
Greek indices run from 0 to 3.

The electric and magnetic fields $\vec E$ and $\vec B$ are expressed
in terms of the potentials by

\begin{equation}
\vec E = - \frac {1} {c}\frac {\partial \vec A}{\partial t} - \vec
\nabla \varphi\, , \qquad \vec B = \vec \nabla \times \vec A\, ,
\end{equation}
so that the homogeneous Maxwell's equations

\begin{equation}
\vec \nabla \times \vec E = - \frac {1} {c}\frac {\partial \vec
B}{\partial t}\, , \qquad \vec \nabla \cdot \vec B = 0\, ,
\end{equation}
are satisfied identically. Employing the tensor $F^{\mu\nu}$
equations (7) are written in a manifestly covariant form as

\begin{equation}
\partial^{\alpha}F^{\beta\gamma} + \partial^{\beta}F^{\gamma\alpha} + \partial^{\gamma}F^{\alpha\beta}=
0\, ,
\end{equation}
which is an identity.

The inhomogeneous Maxwell's equations in the vacuum,
\begin{equation}
\vec \nabla \cdot \vec E = \rho \, , \qquad \vec \nabla \times \vec
B = \frac {1}{c}\vec J + \frac {1}{c}\frac {\partial \vec
E}{\partial t}\, ,
\end{equation}
where $\rho$ and $\vec J$ are the charge and current density, are
written in a manifestly covariant form as\footnote [1] {Expressing
$F^{\mu\nu}$ in terms of $\vec E$ and $\vec B$, one has

$$
F^{10} = E_x\, , F^{20} = E_y\, ,F^{30} = E_z\, ,F^{13} = B_y\,
,F^{21} = B_z\, ,F^{32} \equiv B_x\, ;
$$
for the corresponding dual tensor $\widetilde {F}^{\mu\nu} \equiv
\frac {1}{2} \varepsilon ^{\mu\nu\lambda\rho}F_{\lambda\rho}$, where
$\varepsilon ^{\mu\nu\lambda\rho}$ is the completely antisymmetric
Levi--Civita symbol, one has
$$
\widetilde {F}^{01} = B_x\, , \widetilde {F}^{02} = B_y\,
,\widetilde {F}^{03} = B_z\, ,\widetilde {F}^{13} = E_y\,
,\widetilde {F}^{21} = E_z\, ,\widetilde {F}^{32} \equiv E_x\, .
$$

The Levi--Civita symbol $\varepsilon ^{\mu\nu\lambda\rho}$ is
defined as follows: it equals -1 if the indices form an even
permutation of 0, 1, 2, 3; it equals +1 if the indices form an odd
permutation of 0, 1, 2, 3, and equals 0 otherwise. }

\begin{equation}
\partial_\mu F^{\mu\nu} = \frac {1}{c}J^\nu \, , \qquad J^\nu =
(\rho c,\vec J)\, .
\end{equation}
Expressed in terms of the potentials, eqs. (10) read

\begin{equation}
(\partial_\mu \partial^\mu)A^\nu - \partial ^\nu (\partial_\mu
A^\mu) = \frac {1}{c}J^\nu \, .
\end{equation}
Choosing the Lorenz gauge, $\partial_\mu A^\mu = 0$, the simplified
(post-gauge) equations for the potentials are obtained

\begin{equation}
\Box A^\nu = \frac {1}{c}J^\nu \, , \qquad \Box \equiv \partial_\mu
\partial^\mu = \frac {1}{c^2}\frac {\partial ^2}{\partial t^2} -\vec
\nabla ^2\, .
\end{equation}
Recall that $\varphi$ and $\vec A$ must satisfy the simplified
equations (12) but also the Lorenz gauge; for the retarded
potentials, the last condition is fulfilled if and only if the
continuity equation is satisfied (see, e.g., [7]). The potentials
$\varphi$ and $\vec A$ that satisfy the simplified equations (12)
{\it and} the Lorenz gauge, satisfy also the basic (pre-gauge)
inhomogeneous Maxwell's equations in the vacuum (11).

An analogous remark is also valid if an arbitrary gauge is
introduced into eqs. (11): $\varphi$ and $\vec A$ must satisfy the
corresponding simplified (post-gauge) equations but also the chosen
gauge; for an explicit retarded solution of the simplified
equations, the last condition is fulfilled if and only if the
continuity equation is satisfied [8].

Recall that eqs. (11) are covariant {\it only} by taking that the
foursomes $(\varphi, A_x, A_y, A_z)$ and $(\rho c, J_x, J_y, J_z)$
are four-vectors $A^\mu$ and $J^\mu$ by definition [9]. It should be
stressed that $A^\mu$ is a four-vector by definition, regardless of
whether a covariant or non-covariant gauge is employed in its
calculation in the original reference frame. Clearly, only choosing
a covariant gauge yields the covariant simplified (post-gauge)
equations for the potentials. But, as $A^\mu$ is a four-vector by
definition, the basic (pre-gauge) Maxwell's equations (8) and (11)
are always covariant, employing a non-covariant or covariant gauge
in the original frame. [In addition to satisfying the simplified
equations, the potentials must be consistent with the chosen gauge,
thereby satisfying \emph{also} the basic (pre-gauge) eqs. (11).]

For example, one can use the Coulomb--gauge four--potential
$A^\mu_{\mbox{\tiny C}} = (\varphi_{\mbox{\tiny C}}, \vec
A_{\mbox{\tiny C}})$ in a frame $S$ and Lorentz--transform it to
another frame $S'$ to obtain $A'^\mu_{\mbox{\tiny C}} =
(\varphi'_{\mbox{\tiny C}}, \vec A'_{\mbox{\tiny C}})$. Since the
Coulomb gauge condition $\vec\nabla \cdot \vec A_{\mbox{\tiny C}} =
0$ is not covariant, $\vec\nabla' \cdot \vec A'_{\mbox{\tiny C}}$
does not vanish. But, $\vec E$ and $\vec B$ are gauge invariant, so
one obtains the same $\vec E$ and $\vec B$ in the original $S$ frame
either using the Coulomb gauge or a covariant gauge such as the
Lorenz gauge, $\partial_\mu A^\mu_{\mbox{\tiny L}} = 0$, see, e.g.,
[8]. Now since the field equations written in terms of the
potentials do obey a covariant set of equations, eqs. (8) and (11),
one obtains the same $\vec E'$ and $\vec B'$ in $S'$, regardless of
which gauge is used in the $S$ frame (the gauge employed need not be
covariant). Shortly, while $A^\mu_{\mbox{\tiny C}} \neq
A^\mu_{\mbox{\tiny L}}$, due to the gauge invariance
$F^{\mu\nu}_{\mbox{\tiny C}} = F^{\mu\nu}_{\mbox{\tiny L}}$ and thus
$F'^{\mu\nu}_{\mbox{\tiny C}} = F'^{\mu\nu}_{\mbox{\tiny L}}$. A
telling example is found in Reference [10].

It seems that this subtle point, that the basic Maxwell's equations
(8) and (11) are always covariant due to the fact that $A^\mu$ and
$J^\mu$ are four-vectors by definition, regardless of covariance of
the chosen gauge, is not sufficiently emphasized in classic texts
[4-7]. M{\o}ller's ([4], Section 5.3) and Jackson's ([6], Section
11.9) introduction of $A^\mu$ can be misleading because it is done
using the example of the Lorenz-gauge potentials. (Unfortunately,
the same remark is applicable to Reference [9].) On the other hand,
Griffiths ([7], Subsection 12.3.5) and Landau and Lifshitz ([5],
Sections 16 and 18) seem to imply clearly that the potentials need
not be in the Lorenz gauge to form a four-vector. However, in a
subsequent, somewhat unfortunate footnote, Griffiths states:
``Incidentally, the Coulomb gauge is \emph{bad}, from the point of
view of relativity, because its defining condition, $\bnabla \cdot
\bf A = 0$ is destroyed by Lorentz transformation. To restore this
condition, it is necessary to perform an appropriate gauge
transformation every time you go to a new inertial system, in
\emph{addition} to the Lorentz transformation itself. In this sense,
$A^\mu$ is not a true 4-vector, in the Coulomb gauge.'' ([7], p
574.) This assertion is fundamentally misleading; it erroneously
subordinates the geometric identity of the four-vector potential to
the non-covariance of a specific gauge constraint. By conflating the
transformation properties of the field with the persistence of a
gauge condition, this treatment obscures the fact that $A^\mu$
remains a four-vector regardless of the gauge in use. Also, Landau
and  Lifshitz [5], concerning their eq. (18.1), ``$A'_k = A_k -
\frac {\partial f}{\partial x^k}$,'' fail to emphasize that their
$A_k$ and $A'_k$ are four-vectors by definition--a prerequisite for
the covariance of the basic equations (8) and (11)--regardless of
whether the gauge function $f$ is a Lorentz scalar. This omission
facilitates misinterpretations regarding the four-vector status of
the potential in non-covariant gauges.

Finally, we present three four-vectors frequently used in [1].

First, Pal defines a dimensionless velocity four-vector

\begin{equation}
u^\mu = \gamma\{1, \vec v/c\}\, .
\end{equation}
where $\vec v$ is ordinary velocity three-vector and $\gamma \equiv
(1 - v^2/c^2)^{-1/2}$. He also defines what he calls the electric
field four-vector and magnetic field four-vector, $E^\mu$ and
$B^\mu$, respectively, by eqs. (16) of [1]

\begin{equation}
E^\mu = F^{\mu\nu}u_\nu\, .
\end{equation}

\begin{equation}
B^\mu = -\frac{1}{2}\varepsilon^{\mu\nu\lambda\rho}u_\nu F_{\lambda
\rho} = - \widetilde {F}^{\mu\nu} u_\nu\, .
\end{equation}
Using eqs. (13) - (15) one obtains

\begin{equation}
E^\mu = \gamma(\vec E\cdot \vec v/c, \vec E + \vec v \times \vec
B/c)\, ,
\end{equation}

\begin{equation}
B^\mu = \gamma(\vec B\cdot \vec v/c, \vec B - \vec v \times \vec
E/c)\, .
\end{equation}

One can verify that the electromagnetic field tensor $F^{\mu\nu}$
can be reconstructed from $E^\mu$ and $B^\mu$ as follows (a fact
long established in the literature, see. e.g., [4], Section 7.5)

\begin{equation}
F^{\mu\nu} = E^\mu u^\nu - E^\nu u^\mu +
\varepsilon^{\mu\nu\lambda\rho} B_{\lambda} u_\rho \, .
\end{equation}
The last equation is essential for Pal's covariant extension of
electrodynamics to isotropic media [1].

\section{Covariant electrodynamics in isotropic media revisited}

\subsection{Comments on Pal's original paper [1]}
In Section 3 of [1], which deals with covariant electrodynamics in a
moving isotropic medium, Pal states: ``We define the three--vectors
$\vec D$ and $\vec H$ whose sources are the free charges and
currents, and write the inhomogeneous Maxwell equation in terms of
them. These vectors are assumed to be linearly related to the
electric and magnetic fields $\vec E$ and $\vec B$:

\begin{equation}
\vec D = \epsilon \vec E\, , \qquad  \vec H = \frac {1} {\mu} \vec
B\, .\mbox{''}
\end{equation}

However, in the familiar Minkowski's electrodynamics of moving
media, which is the electrodynamics discussed by Pal, $\vec D$ and
$\vec H$ are defined by

\begin{equation}
\vec D  \equiv \vec E + \vec P \, , \qquad \vec H \equiv \vec B -
\vec M\, ,
\end{equation}
where $\vec P$ and $\vec M$ are the polarization and magnetization
of a moving medium, respectively, and the inhomogeneous Maxwell's
equations in the moving medium read

\begin{equation}
\vec \nabla \cdot \vec D = \rho_{\mbox{\tiny free}}\, , \qquad  \vec
\nabla \times \vec H = \frac {1} {c}\vec {J}_{\mbox{\tiny free}} +
\frac {1} {c}\frac {\partial \vec D} {\partial t}\, .
\end{equation}
Now since

\begin{equation}
\vec \nabla \cdot \vec P = - \rho_{\mbox{\tiny bound}}\, , \qquad
\vec \nabla \times \vec M + \frac {1} {c}\frac {\partial \vec P}
{\partial t}= \frac {1} {c}\vec {J}_{\mbox{\tiny bound}} \, ,
\end{equation}
the sources of $\vec D$ and $\vec H$ are both the free and bound
charges and currents. (Recall that $\vec \nabla \cdot \vec H = -
\vec \nabla \cdot \vec M$; also, in electrostatics $\vec \nabla
\times  \vec D = \vec \nabla \times  \vec P$.)

Neither the physical fields $\vec P$ and $\vec M$ nor auxiliary
fields $\vec D$ and $\vec H$\footnote [2] {It is obvious that $\vec
D  \equiv \vec E + \vec P$ as the sum of two distinct physical
fields must be an auxiliary quantity, and analogously for $\vec H
\equiv \vec B - \vec M$.} appear in Pal's fundamental equations of
his electrodynamics of moving isotropic media. [The only exception
are eqs. (19) (eqs. (12) of [1]), which are not assumptions but
consequences of definitions (20) and empirical relations $\vec P =
(\epsilon - 1) \vec E$ and $\vec M = (\mu - 1) \vec B/\mu$, which
are valid for an isotropic medium at rest.] What Pal presents as
`constitutive relations,' in a moving medium, which (according to
Pal) `are the definitions ascribed to Minkowski,' eqs. (24) of [1],
are nothing but {\it identities}

\begin{equation}
\epsilon E^\mu = \epsilon F^{\mu\nu}u_\nu\, , \qquad H^\mu = \frac
{1}{\mu}B^\mu
\end{equation}
since Pal {\it defines} $E^\mu$ as $F^{\mu\nu}u_\nu$, and $H^\mu$ as
$B^\mu/\mu$, cf. eqs. (16a) and (22b) of [1].

On the other hand, the correct constitutive equations for a moving
isotropic medium, deduced by Minkowski in 1908, read

\begin{equation}
\fl \vec D + \vec v \times \vec H/c = \epsilon (\vec E + \vec v
\times \vec B/c)\, , \quad \vec H - \vec v \times \vec D/c = (\vec B
- \vec v \times \vec E/c)/\mu\, ,
\end{equation}
where $\vec v$ is the velocity of the moving medium [4,7,11,12].

As is well known, {\it defining} a rank-2 antisymmetric tensor
$G^{\mu\nu}$ by

\begin{equation}
G^{10} \equiv D_x\, , G^{20} \equiv D_y\, ,G^{30} \equiv D_z\,
,G^{13} \equiv H_y\, ,G^{21} \equiv H_z\, ,G^{32} \equiv H_x\, ,
\end{equation}
makes it possible to write the inhomogeneous Maxwell's equations
(21) in a manifestly covariant form

\begin{equation}
\partial_\mu G^{\mu\nu} = \frac{1}{c}
J^\nu_{\mbox{\tiny free}}\, , \qquad J^\nu_{\mbox{\tiny free}} =
(\rho_{\mbox{\tiny free}}c, \vec J_{\mbox{\tiny free}})\, ,
\end{equation}
in the general case of a moving anisotropic medium. Introducing
$D^\mu \equiv G^{\mu\nu}u_\nu = \gamma(\vec D\cdot \vec v/c, \vec D
+ \vec v \times \vec H/c)$, where $G^{\mu\nu}$ is given by eq. (25),
and employing eq. (16), the first eq. (24) is recast into the
explicitly covariant form $D^\mu = \epsilon E^\mu$, whereas Pal's
original {\it definition} of $D^\mu$ as $\epsilon E^\mu$, eq. (22a)
of [1], is clearly not a constitutive equation, contrary to his
assertion in [1]. Similarly, introducing the tensor $H^{\mu\nu}$
dual to $G^{\mu\nu}$ given by eq. (25),

\begin{equation}
\fl H^{10} \equiv H_x\, , H^{20} \equiv H_y\, ,H^{30} \equiv H_z\,
,H^{13} \equiv - D_y\, ,H^{21} \equiv -D_z\, ,H^{32} \equiv -D_x\, ,
\end{equation}
and $H^\mu \equiv H^{\mu\nu}u_\nu = \gamma(\vec H\cdot \vec v/c,
\vec H - \vec v \times \vec D/c)$, the second eq. (24) is recast
into $H^\mu = B^\mu/\mu$, employing eq. (17). On the other hand,
Pal's original {\it definition} of $H^\mu$ as $B^\mu/\mu$, eq. (22b)
of [1], is not a constitutive equation.

The constitutive equations (24) imply

\begin{equation}
 \vec D = \gamma^2\Bigg\{\epsilon \left [\vec E + \frac {\vec v}{c} \times \vec B -
 \left (\vec E \cdot\frac {\vec v}{c}\right)\frac {\vec v}{c}\right]-
 \frac {1}{\mu}\frac {\vec v}{c}\times \left (\vec B - \frac {\vec v}{c}\times \vec E\right)\Bigg\} \,
 ,
\end{equation}

\begin{equation}
 \vec H = \gamma^2\Bigg\{\epsilon \frac {\vec v}{c}  \times \left (\vec E + \frac {\vec v}{c} \times \vec B\right )
 + \frac {1}{\mu}\left [ \vec B - \frac {\vec
v}{c}\times \vec E - \left (\vec B \cdot\frac {\vec
v}{c}\right)\frac {\vec v}{c}\right ] \Bigg\} \,
 ,
\end{equation}
Using eqs. (28) and (29) and definitions (20) one obtains the
constitutive equations for $\vec P$ and $\vec M$

\begin{equation}
 \fl \vec P = \gamma^2\Bigg\{(\epsilon - 1)\left [\vec E + \frac {\vec v}{c} \times \vec B -
 \left (\vec E \cdot\frac {\vec v}{c}\right)\frac {\vec v}{c}\right]
 + \frac {\mu - 1}{\mu}\frac {\vec v}{c}\times \left (\vec B - \frac {\vec
v}{c}\times \vec E\right)\Bigg\} \,
 ,
\end{equation}

\begin{equation}
 \fl \vec M = \gamma^2\Bigg\{\frac {\mu - 1}{\mu} \left [\vec B - \frac {\vec v}{c} \times \vec E -
 \left (\vec B \cdot\frac {\vec v}{c}\right)\frac {\vec
 v}{c}\right]- (\epsilon - 1)
 \frac {\vec v}{c}\times \left (\vec E + \frac {\vec
v}{c}\times \vec B \right)\Bigg\} \,
 .
\end{equation}

Thus components of the tensor $G^{\mu\nu}$ given by eq. (25) are
expressed eventually in terms only of $\vec E$, $\vec B$, $\vec v$,
$\epsilon$ and $\mu$, through eqs. (28) and (29). As can be seen,
the last result can be expressed succinctly, `in its full covariant
glory' by equation,

\begin{equation}
\fl G^{\mu\nu} = \epsilon (E^\mu u^\nu - E^\nu u^\mu) + \frac
{1}{\mu} \varepsilon^{\mu\nu\lambda\rho}B_{\lambda}u_\rho\, ,\quad
E^\mu \equiv F^{\mu\kappa}u_\kappa\, , \quad B^\mu \equiv -\frac
{1}{2}\varepsilon^{\mu\nu\lambda\rho}u_\nu F_{\lambda\rho}\, .
\end{equation}
Note that Pal's $G^{\mu\nu}$ (eq. (21) of [1]) is \emph{by
definition} equal to the expression on the right-hand side of eq.
(32); thus, information about $\vec P$ and $\vec M$ is lost in his
approach.

The validity of eq. (32) is not very obvious, since it is based on
non-obvious eq. (18) (eq. (19) of [1]). Moreover, some important
aspects of Minkowski's electrodynamics remain hidden in Pal's
approach which has swallowed $\vec P$ and $\vec M$.\footnote [3]
{For example, the constitutive equations (24) and definitions (20)
imply
$$
 \vec P = (\epsilon - 1) (\vec E + \vec v
\times \vec B/c) + \vec v \times \vec M/c\, ,
$$
Recall that the term $\vec v \times \vec M/c$ in the above equation
is a consequence of a relativistic effect which has no classical
analogue, that a moving magnetic dipole is an electric dipole (see,
e.g., [12]).} It should be stressed, however, that eq. (32) is a
perfectly correct and useful consequence of the definition (25) and
constitutive equations (24), expressed in a compact form.

\subsection{Errors in  Franklin's Comment [2] and Pal's Reply
[3]} In his Comment [2] on Pal's paper [1], Franklin arrives at a
conclusion that the foursome $(\varphi, A_x,A_y,A_z)$ is not a
four--vector $A^\mu$ in a moving isotropic medium. Consequently,
$F^{\mu\nu} = \partial^\mu A^\nu - \partial^\nu A^\mu$ is not a
tensor and thus $G^{\mu\nu}$ is not a tensor in the moving medium.
Franklin eventually concludes that ``... the covariant extension of
electromagnetism to a polarizable medium proposed by Pal fails.''

Franklin's argument goes as follows.

In the vacuum, the wave equation (12) ``shows $A^\nu$ to be a
relativistic four-vector. Then, $F^{\mu\nu}$ [...] is seen to be a
tensor.'' However, in an isotropic medium {\it at rest} $\varphi$
and $\vec A$ satisfy equations

\begin{equation}
\epsilon \left(\frac {\epsilon\mu}{c^2}\partial^2_t - \vec
\nabla^2\right)\varphi = \rho_{\mbox{\tiny free}}\, ,
\end{equation}

\begin{equation}
\frac{1}{\mu} \left(\frac {\epsilon\mu}{c^2}\partial^2_t - \vec
\nabla^2\right)\vec A = \frac{1}{c}\vec J_{\mbox{\tiny free}}\, .
\end{equation}
as follows from eqs. (6), (19) and (21), under the proviso that a
gauge condition

\begin{equation}
\vec \nabla \cdot \vec A + \frac{\epsilon\mu}{c}\frac{\partial
\varphi}{\partial t} = 0\, ,
\end{equation}
has been employed. Now Franklin argues: ``The wave operator now
acting on $\varphi$ and $\vec A$ [in eqs. (33) and (34)] is no
longer a Lorentz scalar so the four-component quantity $A^\nu =
(\varphi, \vec A)$ is no longer a four-vector.'' I understand that
Franklin means by this that $\varphi$ and $\vec A$ now satisfy
non-covariant equations (33) and (34) and hence the foursome
$(\varphi, \vec A)$ cannot be a four-vector.

However, Franklin's logic is invalid, both in the vacuum and
isotropic medium case. It contradicts the well-established argument
of Minkowski's electrodynamics of moving media, {\it which is the
electrodynamics discussed in} [1].

First, the foursome $(\varphi, A_x,A_y,A_z)$ is a four--vector
$A^\mu = (\varphi, A_x,A_y,A_z)$ {\it by definition}, and only thus
$F^{\mu\nu} \equiv
\partial^\mu A^\nu - \partial^\nu A^\mu$ is a rank--2 tensor. Only
thus the homogeneous Maxwell's equations expressed in terms of the
potentials are written in a manifestly covariant form by eqs. (8).
(see, e.g., [9]).

Second, in his Reply [3], Pal has fully refuted Franklin's criticism
of paper [1], noting that eqs. (33)-(35) apply in the rest frame of
a moving medium, and hence the covariance of these equations  should
not be judged from the equations themselves. Pal recalls that``the
crucial question is whether we can write a covariant set of
equations which reduce to equations [(33) and (34)] in the rest
frame of the medium.'' Introducing the four-velocity $u^\mu  =
\gamma \{1,\vec v/c\}$ of a moving medium in the frame considered,
and expressing the terms in eqs. (33)-(35) through $u^\mu$ in a
covariant way, Pal has shown convincingly that eqs. (33)-(35)
written in their ``full covariant glory'' read

\begin{equation}
\epsilon((\epsilon \mu - 1)u_\mu u_\nu
\partial^\mu\partial^\nu + \Box)u_\lambda A^\lambda = \frac
{1}{c}u_\lambda J_{\mbox{\tiny free}}^\lambda
\end{equation}

\begin{equation}
\frac {1}{\mu}((\epsilon \mu - 1)u_\mu u_\nu
\partial^\mu\partial^\nu + \Box)(\eta^{\mu\nu} - u^\mu u^\nu)A_\nu = \frac
{1}{c}(\eta^{\mu\nu} - u^\mu u^\nu) J_{\mbox{\tiny free}\, \nu}
\end{equation}

\begin{equation}
\partial_\mu A^\mu = (1 - \epsilon\mu)(u \cdot\partial)(u \cdot A)\,
,
\end{equation}
respectively.\footnote [4] {Pal also noted correctly that eq. (36)
can be obtained directly from eqs. (26) and (32), expressing $E^\mu$
and $B^\mu$ in eq. (32) through $A^\mu$, contracting eq. (26) with
$u^\nu$ and employing the gauge condition (38), and similarly for
eq. (37).}

To summarize, Pal has established in [3] that reference [2] contains
a serious error by claiming that the foursome $(\varphi,
A_x,A_y,A_z)$ of electromagnetic potentials is not a four-vector in
a moving isotropic medium and that, consequently, Pal's covariant
extension of electromagnetism to isotropic media fails.

It should be stressed, however, that the last paragraph of [3]
contains a basic error. Arguing that $A^\mu$ is a four-vector in the
context of electrodynamics in the vacuum, Pal asserts: ``The
inhomogeneous field equations, written in terms of the potentials,
are covariant {\it only} if the potentials are assumed to satisfy a
covariant gauge condition. In non--covariant gauges such as the
Coulomb gauge or the axial gauge, the potentials do not obey a
covariant set of equations, as can be verified from any textbook on
classical electromagnetic theory.'' However, this is not true.

As is recalled in the present Section 2, the inhomogeneous Maxwell's
equations written in terms of the potentials, the basic eqs. (11),
are always covariant (taking of course that $A^\mu$ and $J^\mu$ are
four--vectors by definition [9]), regardless of whether a covariant
gauge is used or not in the original frame; on the other hand, only
choosing a covariant gauge yields the covariant simplified
(post-gauge) equations for the potentials. Thus, in non-covariant
gauges, the potentials obey a non-covariant set of equations [the
simplified (post-gauge) ones], but also a covariant set of equations
[the basic (pre-gauge) ones], which are valid in any gauge. However,
correcting Pal's error reinstates the fundamental validity of his
rebuttal to Franklin.

An analogous remark applies to Minkowski's electrodynamics of moving
isotropic media: eqs. (8) and (26) are the corresponding covariant
set of equations for the potentials, expressing $G^{\mu\nu}$ by eq.
(32), even if a non--covariant gauge is employed. The
four--potential $A^\mu_{\mbox{\tiny B}}$, calculated in a
non-covariant (``bad'') gauge B, and the four--potential
$A^\mu_{\mbox{\tiny G}}$, calculated in a covariant (``good'') gauge
G, by solving the corresponding simplified equations for the
potentials, non-covariant and covariant, respectively,\footnote [5]
{The key question in Franklin's argument in [2] is whether the gauge
condition (35), valid in the rest frame of an isotropic medium, can
be expressed in a covariant form, valid in an arbitrary inertial
frame, since the basic equations (6), (19) and (21) already belong
to a covariant set of equations, as discussed above. Pal has shown
convincingly that eq. (38) is a covariant generalization of eq. (35)
to the moving medium and thus the simplified equations for the
potentials (33) and (34) must have covariant generalizations too, as
Pal has also demonstrated successfully by eqs. (36) and (37),
respectively.

Clearly, replacing the covariant gauge (35) with a non-covariant one
would yield non-covariant simplified equations for the potentials,
but $A^\mu$ would still be a four-vector which obeys covariant
equations (8) and (26).} yield the same $\vec E$ and $\vec B$ fields
in the original frame due to the gauge invariance, and also in any
other inertial frame, by Lorentz--transforming $A^\mu_{\mbox{\tiny
B}}$ or $A^\mu_{\mbox{\tiny G}}$. (Shortly, while
$A^\mu_{\mbox{\tiny B}}\neq A^\mu_{\mbox{\tiny G}}$, due to the
gauge invariance $F^{\mu\nu}_{\mbox{\tiny B}}=
F^{\mu\nu}_{\mbox{\tiny G}}$, and thus $F'^{\mu\nu}_{\mbox{\tiny
B}}= F'^{\mu\nu}_{\mbox{\tiny G}}$.) Both four--potentials are true
four--vectors from the point of view of relativity. But, ``the
inhomogeneous field equations written in terms of the potentials,''
eqs. (26) via eq. (32), are always covariant, whereas the simplified
equations for the potentials are covariant only in a covariant
gauge.

\section*{Acknowledgment}
I thank Vladimir Hnizdo for illuminating correspondence.

\section*{Conflicts of Interest}
The author declares no conflicts of interest.

\section*{Data Availability Statement}
No new data were generated or analyzed in support of this research.

\section*{Funding}
This research received no external funding.


\Bibliography{99}

\bibitem{P2} Pal, P. B.: Covariant formulation of electrodynamics in isotropic
media. Eur. J. Phys. 43, 015204 (2022)

\bibitem{F} Franklin, J.: Comment on 'Covariant formulation of electrodynamics in isotropic
media'. Eur. J. Phys. 45, 028002 (2024)

\bibitem{P4} Pal, P. B.: Reply to Comment on 'Covariant formulation of electrodynamics in isotropic
media'. Eur. J. Phys. 45, 028001 (2024)

\bibitem{CM} M{\o}ller, C.: The Theory of Relativity. 2nd edn, Clarendon,
Oxford (1972)

\bibitem{LL} Landau, L. D., Lifshitz, E. M.: The Classical Theory
of Fields. 4th rev. edn, Butterworth-Heinemann, Oxford (1996)

\bibitem{JDJ} Jackson, J. D.: Classical Electrodynamics. 3rd edn, Wiley, New
York (1999)

\bibitem{DJG} Griffiths, D. J.: Introduction to Electrodynamics. 5th edn, Cambridge University Press,
Cambridge (2024)

\bibitem{DVR1} Red\v zi\' c, D. V.: A general gauge for the electromagnetic potentials and
the continuity equation. Eur. J. Phys. 37, 065202 (2016)

\bibitem{DVR2} Red\v zi\' c, D. V.: Are Maxwell's equations Lorentz--covariant?.
Eur. J. Phys. 38, 015602 (2017); Corrigendum. Eur. J. Phys. 38,
039501 (2017)

\bibitem{VH} Hnizdo, V.: Potentials of a uniformly moving point charge in the Coulomb gauge. Eur. J. Phys. 25,
351-360 (2004)


\bibitem{HM} Minkowski, H.: Die Grundgleichungen f\"{u}r die
elektromagnetischen Vorg\"{a}nge in bewegten K\"{o}rpern. G\"ott.
Nachr. 53-111 (1908). Reprinted in Minkowski, H.: Gesammelte
Abhandlungen, vol 2, 352-404 Chelsea, New York (1967)


\bibitem{ROSS} Rosser, W. G. V.: An Introduction to the Theory
of Relativity Butterworth, London (1964)

\endbib
\end{document}